\documentclass[aps,prb,twocolumn,superscriptaddress]{revtex4}
\usepackage{graphicx}
\usepackage{calc}
\usepackage{bm}
\begin{document}
\title{Quantum phase transitions in two-dimensional systems}
\date{February 4, 2004}
\author{E.L. Shangina and V.T. Dolgopolov}
\affiliation{ Institute of Solid State Physics RAS, 142432
Chernogolovka, Moscow distr., Russian Federation}

\begin{abstract}
Experimental data on quantum phase transitions in two-dimensional
systems (superconductor-insulator, metal-insulator, and
transitions under conditions of integer quantum Hall effect) are
critically analyzed.
\end{abstract}

 \maketitle
\section{Introduction}

Currently there are quite a few reviews and even books (see, e.g.,
Refs.(\cite{sachdev, sachdev1, sachdev2, lavagna, vojta, schakel,
sachdev3,  schakel1, vojta1}) ), devoted to quantum phase
transitions, where the main focus is on theoretical ideas, while
experimental data are used for illustrative purposes only. The
goal of the present review is to give a critical analysis of the
experimental data on quantum phase transitions  in two-dimensional
systems, which is aimed at revealing reliable established facts,
formulating directions for future research and determining
unresolved problems.

It is convenient to start the explanation of the nature of quantum
phase transitions from continuous phase transitions, i.e., those
not having a stationary coexistence of the two distinct phases
(and, therefore, not having stationary phase boundaries either).
Thereby, at the continuous phase transition point the system as a
whole is changing its phase state. This change of the phase state
is brought into relation with an order parameter which is finite
in one of the phases and permanently becomes zero at the
transition point. Finding an appropriate  order parameter for some
particular phase transition  often presents a  nontrivial problem
in itself. After transition the system becomes stationary and
homogeneously disordered one. Therefore, for all continuous
transitions, as the transition point is approached, diverge both
the duration $\tau_c$ and characteristic size  $r_c$ of the
fluctuations of the order parameter.

The class of continuous phase transitions includes continuous
thermodynamic phase transitions, (e.g., second order phase
transitions ) characterized by singularities in the temperature
derivatives of thermodynamic potentials. The latter are caused by
thermal fluctuations in the system. The divergence of the density
fluctuations size in carbon dioxide ($CO_2$) in the vicinity of a
critical point, corresponding to the continuous  thermodynamic
phase transition, was for the first time experimentally
established in \cite{andrews} by observing a refraction of a
visible light on the density fluctuations.

One can imagine continuous phase transitions to occur at zero
temperature as well. The variation of the system's state is, in
this case, not related to the changing temperature, but to the
variation of a certain external parameter (magnetic field, doping
level, material composition, etc.).  At zero temperature it is, of
course, impossible to register a phase transition through
singularities in the temperature derivatives of thermodynamic
potentials, therefore one should exploit some other properties of
the system, for example, its kinetic characteristics, in order to
find it.

Repeated measurements of a physical quantity with an operator not
commuting with Hamiltonian of the system lead, even for a system
in a stationary state and at arbitrary low temperature, to
different results: the measured quantity experiences quantum
fluctuations. In many measurements one can evaluate dispersion,
and in periodic measurements, the spectral density of states. Both
of these values are determined by the energy of the excited
quantum states of the system. At finite temperature, exceeding the
characteristic energy between stationary states, the main reason
for getting different results in repeated measurements is the
considerable probability of finding the system in different
stationary states, i.e., the thermal fluctuations. For
temperatures about characteristic energy between stationary states
both types  of fluctuations are equally important.

At zero temperature only quantum fluctuations of the order
parameter could drive the phase transition. If tuning parameter
reaches its critical value: $K=K_c$, the system homogeneously
changes the ground state after quantum fluctuation of diverging
size and zero frequency, thus experiencing a quantum phase
transition.

At first glance one might get the impression that quantum phase
transitions cannot be studied experimentally and do not have
practical importance because of impossibility to reach zero of the
temperature. In reality, in the range of temperatures in which
quantum fluctuations compete with thermal ones, and at values of
the tuning parameter close to the critical one, a behavior of the
system is expected to show special features signaling the
existence of a zero temperature quantum phase transition.

\begin{figure}\vspace{-1in}
\scalebox{0.45}{\includegraphics[clip]{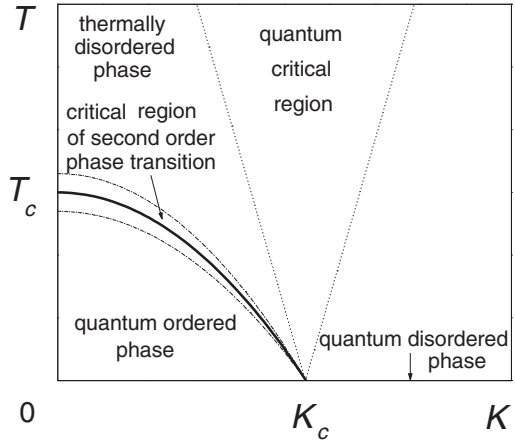}}
\vspace{-1in}\caption{\label{QPT} A diagram of second-order phase
transition. $K=K_c$, and $T=0$ is the quantum phase transition
point. The dotted lines indicate the boundaries of the quantum
critical region.}\vspace{5mm}
\end{figure}

Let us consider, for example, a quantum phase transition located
at the zero-temperature end  of the second order phase transition
curve (Fig.~\ref{QPT}). At finite fixed temperature the increase
of the tuning parameter $K$ leads to a second order phase
transition at the intersection with the solid line in
Fig.~\ref{QPT}. The state of the system is changed through thermal
disordering, i.e., there is a transition from the ordered to the
thermally disordered state.This phase transition is only real
phase transition observed at finite temperature.

At zero temperature, we expect a quantum phase transition from
ordered to disordered state when the external tuning parameter
reaches its critical value $K_c$. As a transition should occur
simultaneously in the whole system, we have to conclude that at
the phase transition point the critical frequency of quantum
fluctuations $ \tau_c ^{-1}$, corresponding to the energy gap
between the ground state of the system and its lowest excited
state, should tend to zero. Simultaneously, the spatial length of
fluctuations $r_c$ (correlation length) should tend to infinity.

Let us return to the case of finite temperature. Because the
critical frequency and correlation length  of quantum fluctuations
are temperature-independent, we can mark, in the $(K,T)$ plane,
the lines on which the critical frequency $\tau_c ^{-1}$ of
quantum fluctuations equals the temperature $kT/\hbar$. These
lines shown by the dotted lines in Fig.~\ref{QPT} border the
so-called quantum critical region, in which the characteristic
length of coherent quantum fluctuations is less than the
correlation length $r_c$ and is restricted by the temperature. At
fixed temperature, a crossover from the thermally disordered  to
the quantum disordered state takes place as this region is
intersected. The observed continuous variation of properties in
the quantum critical region is reminiscent of the quantum phase
transition.

Thereby, the motion parallel to X-axis in Fig.~\ref{QPT}
corresponds, consequently, to a continuous phase transition ,
characterized by its own correlation length of order parameter
fluctuations and attached to the solid line in the figure, and to
subsequent gradual change in the kinetic characteristics of the
system in the quantum critical region. The width of this region
near $K_c$ is dependent on the temperature.

\begin{figure}\vspace{-1in}
\scalebox{0.45}{\includegraphics[clip]{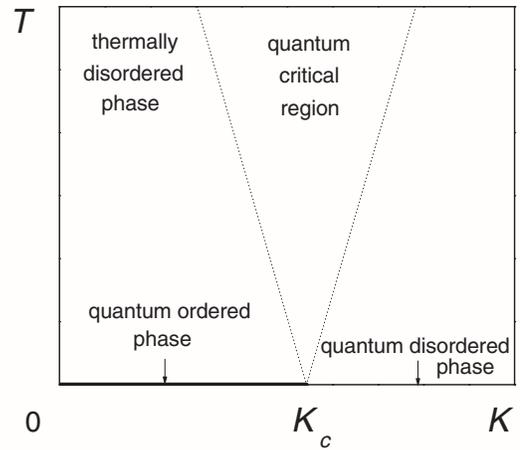}}
\vspace{-1in}\caption{\label{QPT1} Phase diagram of a system
experiencing a quantum phase transition at $K=K_c$, $T=0$. The
dotted lines indicate the boundaries of the quantum critical
region.}\vspace{5mm}
\end{figure}

There are systems in which the existence of the quantum ordered
phase is assumed at zero temperature only. Such a system has a
quantum ordered state at $T=0$, $K<K_c$,experiencing a quantum
phase transition at $K=K_c$. At finite temperatures this system is
disordered (Fig.~\ref{QPT1}). Again, the moving parallel to X-axis
in Fig.~\ref{QPT1} leads to a consequent observation of the
thermally disordered phase, to a gradual change in its properties
towards  those characteristic of quantum disorder and, further,
beyond the boundaries of the quantum critical region, to the
observation of properties peculiar to the quantum disordered
state.

In experiments, the interval of temperatures available for
studying of the transition region properties is principally
restricted from above and below. In the low temperature limit, a
continuous phase transition takes place at $K\approx K_c$. In this
case, when moving along horizontal line in Fig.~\ref{QPT}, the
critical region of the second order phase transition is
inseparable from the quantum critical one. A restriction at high
temperatures is related to the fact that the correlation length
cannot be arbitrarily small and is restricted by characteristic
scales of the problem (coherence length,mean free path, etc.).

In the quantum critical region, a continuous change of
thermodynamic and kinetic properties occurs. Characteristics of
the system are the functions of only one scaling variable $u$,
namely the ratio of the correlation radius  $r_c$ to the
temperature-assigned  dephasing  length  $L_{\Phi}\propto
T^{-1/z}$, where $z$  is the so-called dynamic critical index.
Assuming that in the vicinity of the phase transition point the
correlation radius diverges as $r_c \propto |K-K_c|^{-\nu}$, the
scaling variable can be written in the form
\begin{equation}u=|K-K_c|/T^{1/y},   y=z \nu, \label{eq1}\end{equation}
where $\nu$ is the critical index of the correlation radius. In
other words,  in the quantum critical region one expects that
kinetic characteristics (for example, resistance) will be of the
form
\begin{equation}R=R_0 f(\frac{|K-K_c|}{T^{1/y}}).\label{eq2}\end{equation}
A competition of quantum and classical fluctuations can also be
defined by the ratio of the frequency of critical quantum
fluctuations to the temperature. Using this ratio leads, of
course, to the same scaling parameter $u$  and equation
(\ref{eq2}).

In most experimental papers, an observation of the scaling
relation similar to equation (\ref{eq2}) was considered as a
strong evidence for the quantum phase transition, although in some
publications \cite{sarma, maslov} the remark was made that in
restricted temperature range the occurrence of such a relation can
be accidental.

A behavior typical for quantum phase transition can be observed in
a number of two-dimensional systems. There are the superconductor-
insulator  and metal-insulator transitions as well as transitions
between different quantum states in quantum Hall effect regime.
Below we present the short review of experimental publications in
this field.

\begin{figure}\vspace{-1in}
\scalebox{0.45}{\includegraphics[clip]{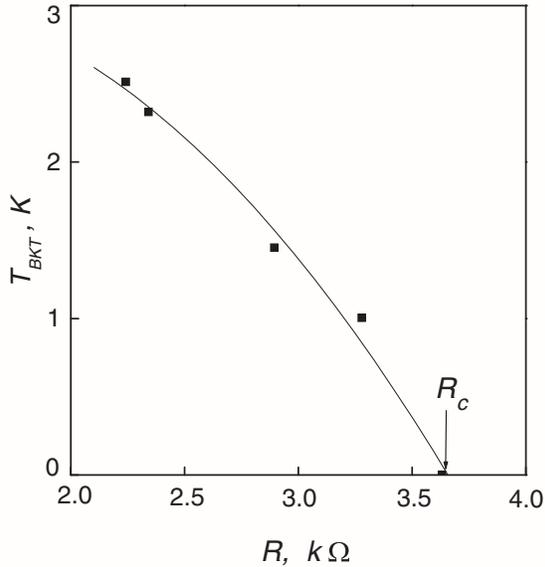}}
\vspace{-1in}\caption{\label{super} Temperature of BKT phase
transition, $T_{BKT}$, as a function of the disorder degree of an
$In/InO_x$ film 10 nm thick. As a measure of disorder in the film,
its resistance per unit surface area at room temperature has been
chosen \cite{hebard1}.}\vspace{5mm}
\end{figure}

\section{Superconductor-insulator phase transition in thin films}

Decreasing one of the sizes of a superconducting sample
$d\ll\lambda$ ($\lambda$ is the penetration length of the magnetic
field into massive sample) lowers the transition temperature to
resistive state. The decrease of this temperature, caused by the
enhanced role of thermal fluctuations in two-dimensional systems,
was first predicted by V L Beresinskii \cite{berez} and
theoretically investigated in Refs
[\cite{kosterlitz,kosterlitz1}]. Since  then a similar continuous
phase transition in thin films of superconductors is known as
Beresinskii-Kosterlitz-Thouless transition (BKT). In the absence
of an external magnetic field the vortices in thin superconducting
film are generated by thermal fluctuations. It is energetically
favorable for vortices with opposite  circulation to form bound
pairs. At low enough temperature $T<T_{BKT}$ , the
'vortex-antivortex' pairs are stable and a film is in the
superconducting state. A temperature increase up to the critical
one, $T=T_{BKT}$, leads to dissociation of 'vortex molecules'
accompanied with the continuous phase transition of a
superconducting film to a resistive state. The temperature of the
BKT phase transition decreases with increasing disorder in the
film. The resistance of the film can be considered as a measure of
its disorder. Dependence of $T_{BKT}$ on the disorder degree is
shown in Fig.~\ref{super}. Films with a resistance smaller than
the critical one, $R_c$, experience BKT transition, but  for
$R>R_c$  the film is still in the resistive state up to the lowest
experimentally reachable temperatures.

\begin{figure}\vspace{-1in}
\scalebox{0.45}{\includegraphics[clip]{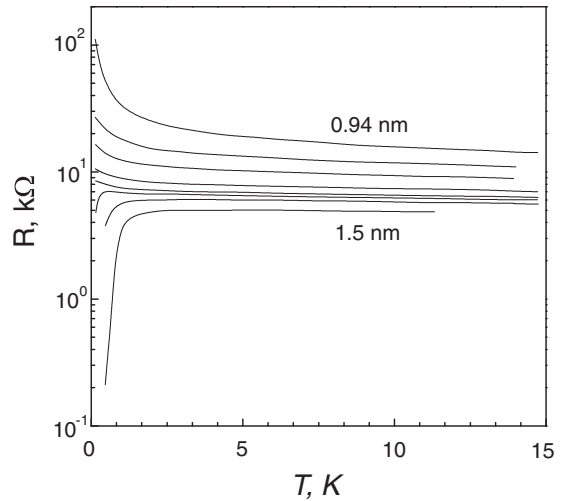}}
\vspace{-1in}\caption{\label{mark} Temperature dependencies of
resistance per unit surface area of amorphous $Bi$ films of
varying thickness $d=0.94\div1.5$ nm, changing with the step of
0.08 nm (according to the data of
Ref.[\cite{markovic}]).}\vspace{5mm}
\end{figure}

BKT phase transition can be driven by the change of another
external parameter, e.g., the film thickness or magnetic field. In
Fig.~\ref{mark}, temperature dependencies of the resistance of
amorphous bismuth films of varying thickness $d=0.94\div1.5$ nm
are presented. The films having a critical thickness $d_c
\simeq1.3$ nm have an approximate temperature-independent value of
resistance $R_c\approx 7$ kOhm. In the low-temperature range,
films having a thickness exceeding the critical one exhibit a
positive derivative $dR/dT>0$  characteristic of metallic
conductivity and, with  further temperature decrease, experience a
transition to the superconducting phase. Films having a thickness
less than the critical one show, however, a negative derivative
$dR/dT<0$. A decrease in the film's thickness down to
$d\approx0.9$ nm increases its resistance up to $\sim10^4$ kOhm.
Therefore, films having a thickness $d<d_c$ show quasi-insulating
properties without experiencing the BKT transition.

\begin{figure}\vspace{-0.5in}
\scalebox{0.45}{\includegraphics[clip]{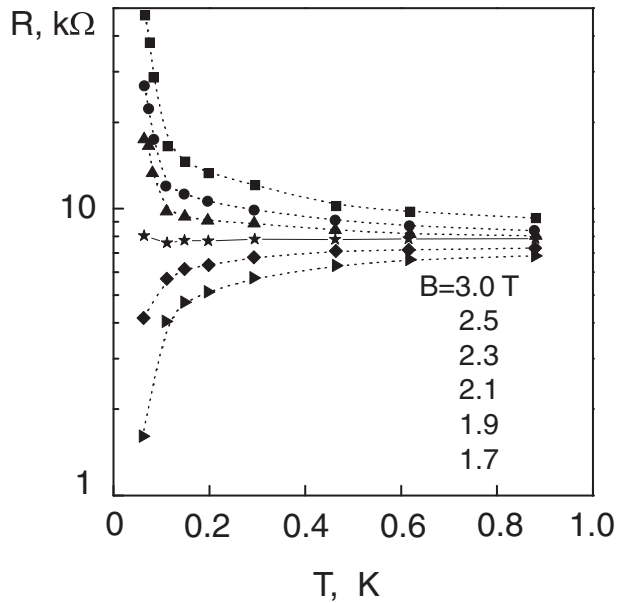}}
\vspace{-0.5in}\caption{\label{isomag} Temperature dependencies of
the resistance of an amorphous $InO$ film with thickness $d=20$
nm, measured at different  magnetic fields $B=1.7\div3.0$ T
(according to the data of Ref.[\cite{gant1}]).}\vspace{5mm}
\end{figure}

The magnetic field has an analogous influence on BKT phase
transition. Isomagnetic curves for the temperature dependence of
resistance of amorphous $InO$ film ($d=20$ nm) in the magnetic
fields  $B=1.7\div3.0$ T, are shown in Fig.~\ref{isomag}. The
temperature-independent value of the film's resistance $R_c\approx
8$ kOhm corresponds to the critical magnetic field  $B_c\approx
2.1$ T. For magnetic fields lower than the critical one, the film
shows a positive temperature coefficient of resistance $dR/dT>0$
typical for metallic state. At  lowering the temperature the film
with metallic conductivity demonstrates the transition into
superconducting state (see Fig.~\ref{isomag}). For magnetic fields
larger than the critical one, the film exhibits quasi-dielectric
properties with a negative derivative $dR/dT<0$ and no signs of
its transition to the superconducting state can be observed down
to the temperature $T=0.035$ K.

Therefore, the available experimental data allow us to conclude
that when an external parameter (film thickness, degree of
disorder, magnetic field) reaches its critical value at zero
temperature, the ground state of the film changes in a fundamental
way: from the superconducting state to the insulating state. In
other words, a thin film undergoes the quantum phase transition
from superconducting to insulating state at $K=K_c$. The
pioneering paper \cite{fisher}  provides a theoretical ground for
quantum superconductor-insulator phase transition in thin films.

A quantum superconductor-insulator transition  can be considered
as the quantum analogue of continuous BKT phase transition. At
absolute zero of the temperature, the vortices  arising in a thin
film due to quantum fluctuations are localized ('pinned' by the
defects) and form so called 'vortex glass'. Strengthening  the
external magnetic field increases the concentration of vortices
with orientation corresponding to that of the field. The growth of
the disorder degree in the film (or decrease of its thickness)
also increases the concentration of the vortices. When a
concentration of vortices approaches its critical value, the
localization length of vortices diverges as a function of
$|K-K_c|$. Finally, at the critical value of the external
parameter $K=K_c$, the vortices delocalize. As shown in
Ref.[\cite{fisher}], delocalization of vortices is necessarily
accompanied by the localization of Cooper pairs, thus leading to
the formation of the so-called 'Cooper pair glass'. Such
'complementarity' in the behavior of two boson systems is due to
the duality of their Hamiltonians in a two-dimensional film
\cite{fisher}. At zero temperature, a localization of Cooper pairs
means the transition from superconducting to insulating state. The
metallic state, in which Cooper pairs and vortices move
diffusively with a finite resistance, is an intermidiate one  in
between the insulating and superconducting states at absolute zero
of the temperature.

\begin{figure}\vspace{-0.8in}
\scalebox{0.45}{\includegraphics[clip]{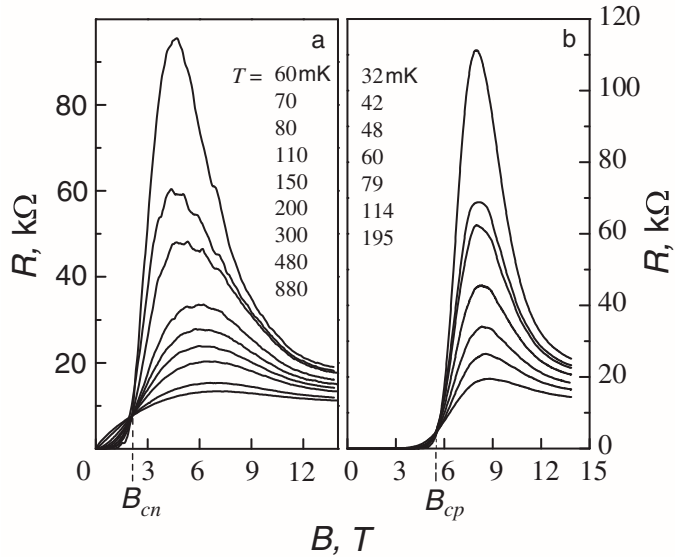}}
\vspace{-0.8in}\caption{\label{super1} Transition of amorphous
$InO$ film with thickness $d=20$ nm from the superconductive to
the resistive state \cite{gant}. The film resistance is shown as a
function of the external magnetic field having (a) normal, and (b)
parallel orientation in the temperature range $T=32\div880$ mK.
Dashed lines mark the critical values of the magnetic field,
corresponding to the quantum superconductor-insulator phase
transition.}\vspace{5mm}
\end{figure}

\begin{figure}\vspace{-0.8in}
\scalebox{0.45}{\includegraphics[clip]{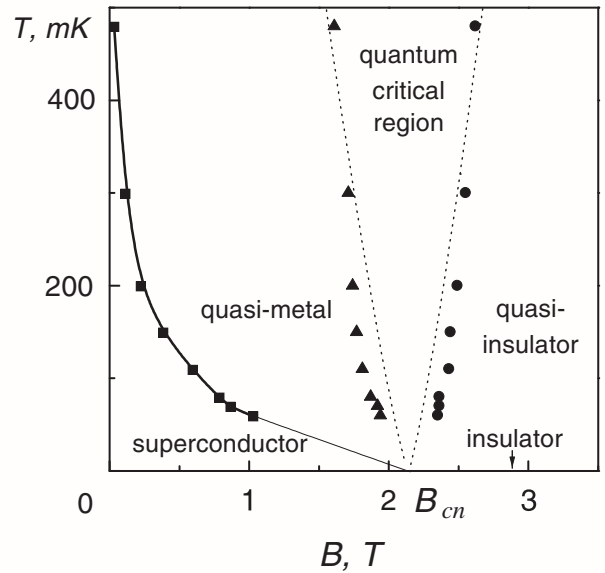}}
\vspace{-0.8in}\caption{\label{super2} Diagram of BKT phase
transition in an amorphous $InO$ film with thickness $d=20$ nm in
$(B,T)$ coordinates (according to the data of Ref. [\cite{gant}]).
The magnetic field is oriented normally to the film surface. The
solid line marks the boundary between the phases, and the dotted
lines mark the boundaries of the quantum critical
region.}\vspace{8mm}
\end{figure}

As an example, let us consider magnetic-field-induced transition
of an amorphous $InO$ film with thickness $d=20$ nm from
superconducting to insulating state \cite{gant}. These films are
two-dimensional for vortices, because the penetration length of
the magnetic field is $\lambda\geq 100$ nm. However, for normal
electrons the  $InO$ film is a bulk sample, because the electron
mean free path  in the film is $l\sim 1$ nm. Experimental
dependencies of the resistance of such a system on the normal to
the film's surface magnetic field are shown in
Fig.~\ref{super1}$^a$ for various temperatures. The
temperature-independent value of the resistance $R_{cn}\approx 8$
kOhm corresponds to the critical magnetic field $B_{cn}\approx
2.2$ T. At a finite temperature, the magnetic field value at which
the film resistance becomes finite corresponds to the transition
to resistive state or BKT transition (see Fig.~\ref{super2}). The
boundary of BKT phase transition in Fig.~\ref{super2} is
determined from the film resistance exceeding 1/1000 of its
maximal value in the resistive state. The boundaries of quantum
critical region at an arbitrary temperature, $B(T)$, can be
determined from the coincidence of resistance measured at the
maximal temperature in the experiment, $T=T_{max}$, with that
corresponding to the temperature  $T$, after normalizing of the
resistance according to $R(B)=
R^{norm}((B-B_{cn})*(T_{max}/T)^{1/z\nu})$. The thus found
boundaries of the quantum critical region in the temperature range
$T=60\div480$ mK are shown in Fig.~\ref{super2} by circles and
triangles.

The phase diagram of BKT phase transition in the normal magnetic
field (see Fig.~\ref{super2}) is qualitatively similar to the
theoretical model, presented in Fig.~\ref{QPT}. There exists,
however, difference in the sign of the second derivative of the
BKT phase transition temperature in the plane $(B,T)$. According
to the model ideas \cite{fisher}, the BKT phase transition
temperature changes as $T_{BKT}\propto(B_c-B)^{0.5}$ with a
negative second derivative (see Fig.[~\ref{QPT}]). Experimentally,
however, this second derivative is positive and does correspond to
$T_{BKT}\propto |B-B_{cn}|^{2.49}$.

Let us now discuss the phase boundaries of quantum critical region
in $(B,T)$ plane. In Refs.[\cite{gant1,gant}], it was shown that
in the quantum critical region  the resistance of amorphous $InO$
films in the normal magnetic field is a function of the scaling
variable $u\propto|B-B_{cn}|T^{-1/z\nu}$ with exponent
$z\nu=1.15\div 1.22$. The value of the product $z \nu$ of critical
indices obtained in Refs.[\cite{gant1,gant}] is not universal. It
was found, e.g., for amorphous $InO_x$ and $MoGe$ films $z
\nu=1.26\div1.31$ \cite{hebard} and $1.27\div1.37$ \cite{yazdani},
correspondingly. In amorphous and granular $In$ films $z
\nu=0.48\pm0.04$ and $0.62\pm0.04$ \cite{okuma}; in
$Nd_{2-x}Ce_xCuO_4 (x\sim0.15)$ $z \nu\sim0.5$ \cite{suzuki}, and
in amorphous Bi films $z \nu=0.7\pm0.2$ \cite{markovic,markovic1}.
For transitions in amorphous bismuth films with varying thickness
in a zero or constant normal magnetic field it was found
$z\nu=1.4\pm0.2$ \cite{markovic,markovic1}. In
Refs.[\cite{yazdani,markovic1}], in additional studies of the
electric field scaling  for  $MoGe$ and $Bi$ films the universal
value  $z\approx 1.0$ of the dynamic critical index was obtained.
The  spread in the values  of the product $z \nu$ of indices in
thin films  was explained by the variations in the critical index
$\nu$.

Let us use the value   $z \nu=1.15\div 1.22$ \cite{gant1,gant},
obtained in the studies of temperature scaling, for constructing
the expected boundaries of the quantum critical region for quantum
phase transition in the normal magnetic field. The corresponding
boundaries  are shown in Fig.~\ref{super2} by dotted lines. As
seen from the figure, the experimentally derived boundaries does
not quite correspond to the theoretically expected one.

\begin{figure}\vspace{-0.8in}
\scalebox{0.45}{\includegraphics[clip]{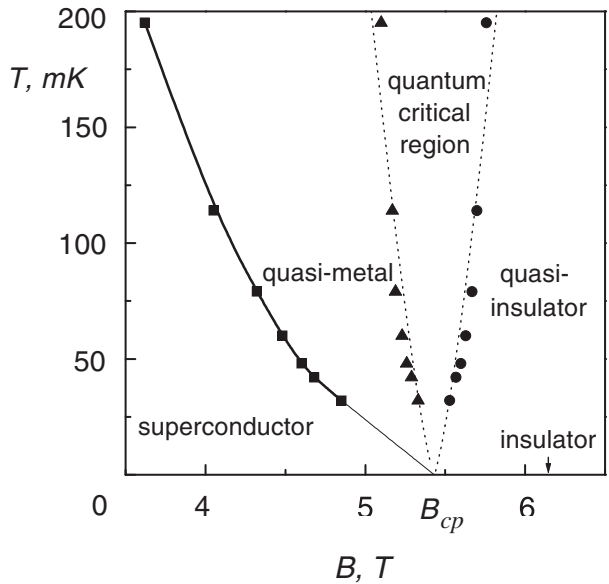}}
\vspace{-0.8in}\caption{\label{super4} Diagram of continuous phase
transition in an amorphous $InO$ film with thickness $d=20$ nm
from superconducting to resistive state in $(B,T)$ coordinates
(according to the data of Ref. [\cite{gant}]). The magnetic field
is oriented parallel to the film surface. The solid line marks the
boundary between the phases, and the dotted lines mark the
boundaries of the quantum critical region.}\vspace{5mm}
\end{figure}

Besides the phase transition in amorphous $InO$ films in the
normal magnetic field, Gantmakher et al. \cite{gant}  studied a
transition from the superconducting to resistive state in the
magnetic field oriented parallel to the film surface. The
experimental dependencies of the $InO$  film resistance on
parallel magnetic field at different temperatures are shown in
Fig.~\ref{super1}b.  The isotherms $R(B)$ intersect at the
critical value  $B_{cp}\approx 5.4$ T of the magnetic field. The
observed crossing of isotherms looks very much like evidence of a
quantum phase transition between superconducting and insulating
states of the film in the parallel field  $B=B_{cp}$. In the
critical magnetic field the resistance of the film is temperature
independent and equal to $R_{cp}\approx 5$ kOhm. Using the
experimental data from Fig.~\ref{super1}b, we find the phase
transition boundary and the boundaries of the quantum critical
region in the temperature range $T=32\div195 $ mK by the method
which has been described above for the transition in the normal
magnetic field. The corresponding phase diagram for the case of
parallel magnetic field is shown in Fig.~\ref{super4}.

Phase diagrams of the transition of an $InO$ film from the
superconducting to the insulating state in the parallel and the
normal magnetic fields are strikingly similar. The temperature of
transition to resistive state decreases with increase of magnetic
field irrespective of its orientation. In the parallel magnetic
field , $T_c\propto (B_{cp}-B)^{1.78}$ (Fig.~\ref{super4}) which
is close to $T_{BKT}\propto |B-B_{cn}|^{2.49}$, found in the
normal orientation . In Ref.[\cite{gant}] it was shown that in the
parallel magnetic field scaling exponent is equal to  $z
\nu=1.30$. Knowing the product of critical indices, it is easy to
draw an expected boundary of the quantum critical region (dotted
lines in Fig.~\ref{super4}). As seen from the figure, the
agreement between the expected critical region boundaries and the
experimentally found points is distinctly better for the parallel
field than for the normal one.

Hence, although the experiments on studying the superconductor-
insulator phase transition in two-dimensional objects
qualitatively  confirm theoretical predictions \cite{fisher}, they
also reveal a number of problems. First, a BKT transition boundary
has an unexpected form. Second, the theory developed for the
normal magnetic field and essentially using the fact of the normal
field orientation, is unexpectedly formally suitable for
describing results in the magnetic field parallel to the film
surface.

\section{Phase transitions in the integer quantum Hall effect regime}

It is considered as evident that in the absence of a magnetic
field a two-dimensional electron system is an insulator in
arbitrary chaotic potential \cite{abrahams}. This statement, valid
for the systems in which one can neglect electron-electron
interaction, means that at zero temperature the conductance of a
two-dimensional system  starting from some, generally speaking,
large size is exponentially decreasing with a further increase in
a size of the system. In quantizing magnetic field with
$\omega_c\tau\gg1$, where $\omega_c=eB/m^*$ is the cyclotron
frequency, and  $\tau$ is the momentum relaxation time of
electrons, the ground state of the system depends on the relation
between the field strength and the density  $n_s$ of
two-dimensional electrons, determined by the filling factor
$\nu^*=n_s/n_B$, where $n_B=eB/h$ is the number of magnetic flux
quanta $h/e$ per unit surface.

\begin{figure}\vspace{-0.8in}
\scalebox{0.45}{\includegraphics[clip]{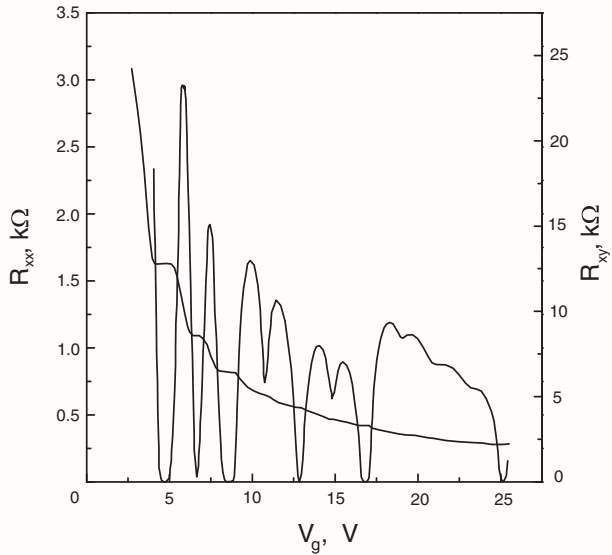}}
\vspace{-0.8in}\caption{\label{klitzing} Discovery of the integer
quantum Hall effect in the two-dimensional electron system in a
silicon MOSFET \cite{klitzing}. For the first time, the horizontal
plateaus in the Hall resistance $R_{xy}$ and corresponding minima
of magnetoresistance $R_{xx}$ at the temperature $T=1.5$ K were
observed. On the horizontal axis, the values of the gate voltage
$V_g$, which changes the concentration $n_s$ of the carriers and,
correspondingly, the filling factor in the constant magnetic field
of 18 T, are plotted.}\vspace{5mm}
\end{figure}

As it was experimentally found in 1980 in a silicon MOSFET
\cite{klitzing},  in the vicinity of the  integer filling factors
the diagonal resistance $R_{xx}$ takes a zero value, whereas the
Hall component $R_{xy}$ shows a set of quantized plateaus. In the
vicinity of the half-integer filling factors, $R_{xx}$ has maxima,
and $R_{xy}$ jumps from one plateau to another
(Fig.~\ref{klitzing}). Such a behavior of the components  of
resistance tensor, known as integer quantum Hall effect (IQHE),
was interpreted as the existence of a number of insulating phases
with zero dissipative conductivity and quantized Hall conductivity
separated by metallic states\cite{pruisken}. Experimentally, it is
hard to proof the exact  $\sigma_{xy}$ quantization, because a
simple inversion of the resistance tensor assumes a uniform
current flow, whereas under conditions of the quantum Hall effect
a significant part of the current flows near the sample edge
\cite{thouless1}.

Consideration of IQHE as consequence of quantum phase transitions
in the strong magnetic field \cite{pruisken, levine} raises two
principal questions: the first about the system behavior at
zeroing of magnetic field, and the second  on the detailed
description of the transition region between quantum plateaus. The
first question was theoretically considered  in
Ref.[\cite{khmelnitskii}], where a chain of quantum phase
transitions with quantized   $\sigma_{xy}$ values was predicted in
the region of classically weak magnetic fields. Although the
proposed picture, known as 'floating of extended states', has a
number of indirect experimental confirmations
\cite{shashkin,jiang,dultz}, there remain doubts about the
possibility of realizing such a chain in samples of reasonable
size at reasonable temperatures \cite{huskestein}. Below we
discuss available experimental information regarding the second
question.

\begin{figure}\vspace{0.1in}
\scalebox{0.35}{\includegraphics[clip]{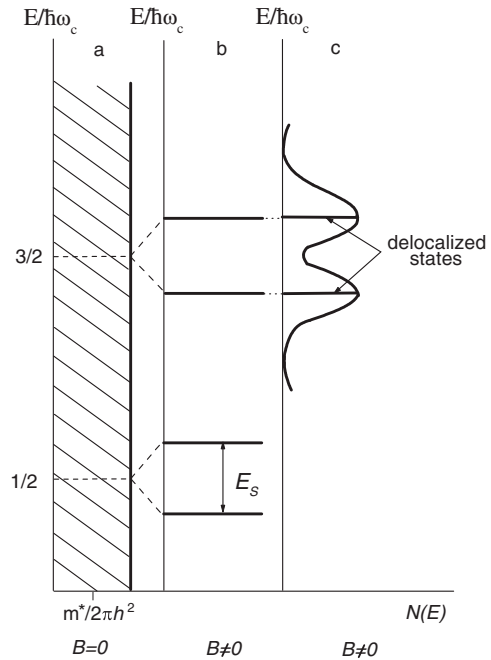}}
\vspace{0.01in}\caption{\label{landau} Dependence of the density
of states of a two-dimensional system of noninteracting electrons
on energy: (a) in the absence of a magnetic field; (b) in the
absence of scattering by chaotic potential, and (c) with a finite
magnetic field and scattering.}\vspace{4mm}
\end{figure}

Let us choose, as an example, a two-dimensional electron system in
the long-period chaotic potential with the characteristic size
$l_0\gg l_B$ in the plane,  where   $l_B$ is the magnetic length
\cite {Iordanskii}. The energy spectrum of an ideal system of
noninteracting electrons in the magnetic field is presented by a
set of delta functions ordered along the energy axis in accordance
with the values of cyclotron energy $\hbar\omega_c$  and the spin
splitting energy $E_s$ (Fig.~\ref{landau}, b). A long-period
chaotic potential causes the inhomogeneous broadening of each of
quantum levels (Fig.~\ref{landau}, c) so that at each level only
one state, corresponding to the percolation threshold, is
delocalized. The other electrons are localized near some extremes
of the chaotic potential. At the tuning of the carrier density or
of the  magnetic field, i.e. at the tuning of filling factor, the
Fermi level sequentially crosses the bands of localized states in
which $\sigma_{xx}=0$,  and $\sigma_{xy}$ takes a quantized value
$i(e^2/h)$ ($i$-is an integer). Transition between insulating
phases with different indices $i$ occurs via the metallic phase
corresponding to the coincidence of the Fermi level  $E_F$ with
the energy of delocalized state $E_i$. The number $i$ of
delocalized states under the Fermi energy, determining the value
of $\sigma_{xy}$, changes by one and the dissipative conductivity
shows a sharp peak.

In the symmetric potential at zero temperature, the condition
$E_F=E_i$ corresponds to the critical filling factor
$\nu^*_{ci}=i-1/2$. When approaching the critical filling factor ,
a localization length of the electrons at the Fermi level tends to
infinity as $\xi \propto |E_F-E_i|^ {-\nu} \propto |\nu^* -
\nu^*_{ci}|^{-\nu}$. At a finite temperature one expects a gradual
change in Hall conductance and a broadening of the dissipative
conductance peaks in the quantum critical region of the transition
between two insulating phases. The scaling parameter is a ratio of
the temperature-assigned coherence length $L_{in}(T)\propto
T^{-p/2}$ (Ref.[ \cite{thouless}]) to the localization length of
carriers at the Fermi-level:
\begin{equation}u={(L_{in}(T)/\xi(\nu^*))}^{1/\nu}\propto|\nu^*
- \nu^*_{ci}|T^{-p/2\nu}.\label{eq3}\end{equation}

In the quantum critical region, the components of the conductance
tensor  $\sigma_{\alpha\beta}$  or conventionally measured in
experiments resistivity tensor  $\rho_{\alpha \beta}$, are
expected to be functions of the scaling parameter  $u$.

The  $m$th order derivatives of kinetic characteristics taken at
critical point depend on the temperature according to a power law

\begin{equation}{(\partial^m\rho(\nu^*)_{\alpha \beta}/\partial \nu^{*m})}_{\nu^*=
\nu^*_{ci}}\propto T^{-mp/2\nu}.\label{eq4}\end{equation}

As follows from Refs.[\cite{pruisken,levine}], the above described
properties are valid for an arbitrary electron system in an
arbitrary chaotic potential, if at each quantum level there exists
one delocalized state with infinite localization length.

\begin{figure}\vspace{-0.8in}
\scalebox{0.45}{\includegraphics[clip]{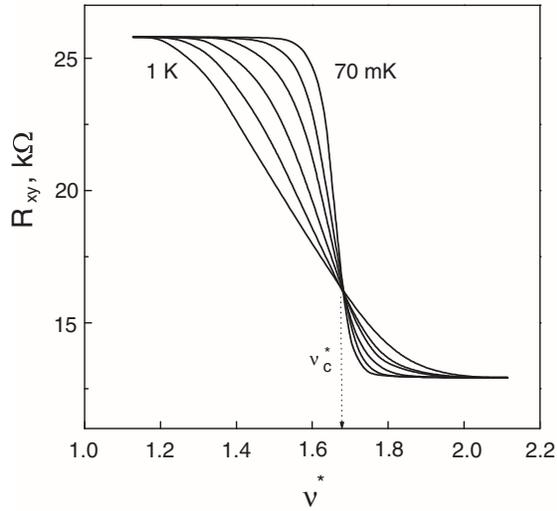}}
\vspace{-0.8in}\caption{\label{dunford} Integer quantum Hall
effect in the two-dimensional hole system $Si/Si_{0.87}Ge_{0.13}$
(Ref.[\cite{dunford}]). Dependencies of the Hall resistance
$R_{xy}$ on the filling factor $\nu^*$ at the temperatures 70,
190, 330, 500, 700 and 1000 mK are shown. The arrow marks the
critical value of the filling factor corresponding to a quantum
phase transition between the states of a Hall insulator with $i$=2
and $i$=1.}\vspace{4mm}
\end{figure}

As an example, let us consider the phase transitions between the
Hall insulators in the two-dimensional hole system
$Si/Si_{0.87}Ge_{0.13}$ \cite{dunford}. Experimental dependencies
of the Hall resistance of such a system on the filling factor are
shown in Fig.~\ref{dunford} for different temperatures. The
temperature-independent value of Hall resistance $R_{xy}\approx
16$ kOhm corresponds to the critical filling factor
$\nu^*_c\approx 1.68$  for a quantum phase transition between Hall
insulators with $i$=1 and  $i$=2. In insulating states  with $i$=1
and $i$=2 Hall resistances $R_{xy}$ are equal to $h/e^2\approx 26$
kOhm and $h/2e^2\approx 13$ kOhm, respectively.

\begin{figure}\vspace{-0.8in}
\scalebox{0.45}{\includegraphics[clip]{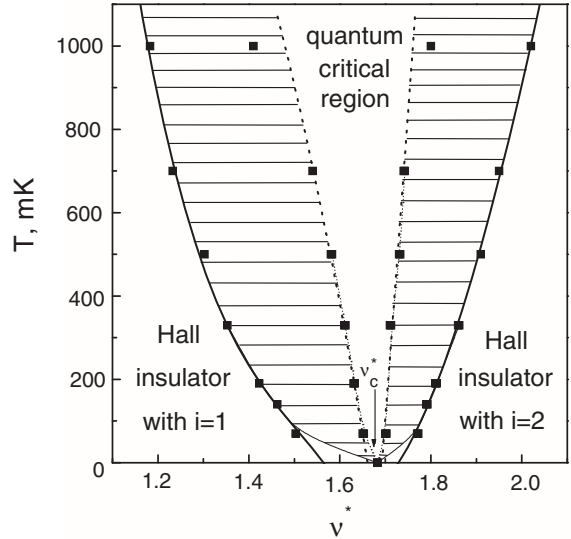}}
\vspace{-0.8in}\caption{\label{phase} Diagram of continuous phase
transition between the states of a Hall insulator with $i$=1 and
$i$=2 through the thermally disordered phase with metallic
conductance (according to the data of Ref.[\cite{dunford}]). Solid
lines are the lines of phase transitions, dashed lines show the
boundaries of the quantum critical region. Horizontal hatching
marks the portion of the phase diagram corresponding to the
thermally disordered phase with metallic conductance.}\vspace{1mm}
\end{figure}

Based on the data presented in Fig.~\ref{dunford} it is possible
to construct a phase diagram similar to those shown in
Fig.~\ref{super2} and Fig.~\ref{super4}. Deviation of the Hall
resistance from the quantized values  $h/ie^2$ corresponds to the
metallic state of the system, which is separated from the
insulating states by the phase boundaries, shown by solid lines in
Fig.~\ref{phase}. The boundaries of quantum critical region
$\nu^*(T)$ can be determined through  coincidence of the Hall
resistance, measured at the maximum temperature of experiment
$T=T_{max}$, with Hall resistance at $T$, after normalizing of the
resistance according to $R_{xy}(\nu^*)=
R_{xy}^{norm}((\nu^*-\nu^*_{c})*(T_{max}/T)^{p/2\nu})$. The thus
-determined boundaries of the quantum critical region are marked
in Fig.~\ref{phase} by rectangles.

Motion along the horizontal line at constant temperature in
Fig.~\ref{phase}  displaces the Fermi level from one band of
localized states to another one while crossing the delocalized
state with infinite localization length.  This motion corresponds
to sequence of phase transitions from the insulating state $i$=1
with Hall conductance $\sigma_{xy}=e^2/h$ to the metallic state,
and then to the new insulating state with  $i$=2.

It was shown in Refs.[ \cite{dunford, dunford1}] that for
$Si/Si_{0.87}Ge_{0.13}$ structures in the quantum critical region,
the scaling relations  (\ref{eq3}) are really fulfilled with
exponent  $k=0.70\pm0.05$ for the transition between Hall
insulators with  $i$=1,$i$=2 and with exponent $k=0.68\pm0.05$
 for the transition between $i$=0 and $i$=1. The analysis of
 experimental data was based on the assumption that all phase
boundaries depicted in Fig.~\ref{phase} will, on decreasing
temperature, shrink into one point as shown by thin solid line and
by dotted line in the figure. This assumption corresponds to the
existence of only one delocalized state with infinite localization
length at each of the quantum levels and can be substantiated only
for noninteracting electrons in the long-period chaotic potential.

It appears natural to extrapolate the phase boundaries to zero
temperature according to the experimentally established law which
is, as seen from Fig.~\ref{phase}, close to the linear one. If
such an extrapolation is correct, one should come to a conclusion
on the finiteness of the band width of metallic state at zero
temperature  and, correspondingly, on the existence of two quantum
phase transitions in the intervals  between insulating phases.
Although the experimental evidence for the finite bandwidth of
metallic state at zero temperature was found in a number of papers
\cite{dolgop,balaban,shahar,arapov}, in none of them was performed
a scaling analysis assuming the existence of two successive
quantum phase transitions.

Following the pioneering work by Wei et al.\cite{wei}, an analysis
of the experimental data assuming one extended state at the
quantum level was done for two-dimensional systems  à $InGaAs/InP$
\cite{wei3,wei2,wei4,pan,scalIn}, $AlGaAs/GaAs$
\cite{wei1,scalGaAs4}, and $GeSi/Ge$ \cite{coleridge}. In all
these works, the exponent was equal to $k=0.42\div0.46$. In Ref.[
\cite{scalIn1}], the value of  $k=0.57$ for transitions between
the Hall insulators with $i$=1 and $i$=0 in the system
$InGaAs/InP$ was obtained. For transitions between the spin-
degenerate insulating states in IQHE, a value approximately two
times less, $k \approx 0.2$ was found \cite{wei3,wei2}. In
experiments involving observation of IQHE in Si-MOSFET
\cite{japan} and two-dimensional $AlGaAs/GaAs$ systems, differing
in the type and concentration of the dopant \cite{koch}, a
dependence of the value of $k$ on the number of the Landau level,
carrier mobility and doping parameters was observed.

The spread in values of the scaling exponent $ê$ is described by
the difference in the mechanisms of inelastic electron scattering
in various systems, which determine the value of the exponent $p$
in the temperature dependence of the coherence length
\cite{dunford1}, or  by transitions belonging to different classes
of universality \cite{wei4}. The principal question on the
bandwidth of the delocalized states, especially in systems with
distinct effects of electron-electron interactions, escaped the
attention of a majority of researchers.

\section{Metal-insulator transition in two-dimensional systems }

Is it possible to observe a metal-insulator transition in a
two-dimensional system in the limit of a zero magnetic field?
Thirty five  years ago there was no doubt about the answer: the
transition is possible , and this is the Mott-Anderson transition.
The publication of the theoretical paper Ref.[ \cite{abrahams}] in
1979 radically changed  the answer to this question. The authors
of Ref.[\cite{abrahams}] employed a scaling approach to an
analysis  of the conductance of systems  in the approximation of
noninteracting carriers. According to the scaling hypothesis, a
logarithmic derivative of dimensionless conductance $g=2\hbar
G/e^2$ with respect to the system size $L$ at zero temperature is
a function only of the conductance itself. For two-dimensional
systems in the absence of spin-orbit interaction, this derivative
is negative in the whole range of values of dimensionless
conductance. This means that with unlimited growing size of the
system its conductance is continuously decreasing, i.e., a
two-dimensional system of infinite size is, at zero temperature,
in an insulating state with zero conductance independent of how
large the initial conductance of the finite system was.
Electron-electron interaction in the 'dirty' limit additionally
enhances the localization of carriers \cite{altshuler}. A theory
of quantum corrections (TQC) to conductance
\cite{abrahams,altshuler,hikami,kawabata,altshuler1,lee,finkel}
that  considers a phenomenon of weak localization and
electron-electron interactions in disordered systems confirmed the
asymptotic form of the scaling function at large values of
conductance.

The following two almost complete decades can be named a time of
triumph of TQC. This theory allowed the explanation and
classification of the experimentally derived low-temperature
anomalies in kinetic effects, in particular, negative
magnetoresistance and logarithmic temperature dependence of the
conductance of 'dirty' semiconductor heterostructures with
two-dimensional electron or hole gas. The first observations of
the TQC-predicted logarithmic dependence of conductance on
temperature in Si-MOSFET were made in Refs.[\cite{tkp1,tkp2}]. The
experimentally  established negative magnetoresistance in
Si-MOSFET  \cite{oms,oms1,oms2,oms3} was also analyzed from the
TQC point of view. Characteristic sizes of the self-crossing
trajectories, phase relaxation time of the electron wave function
due to electron-electron  and electron-phonon collisions, and the
electron-electron coupling constant in the diffusive channel were
determined. Mechanisms  of the energy relaxation of electrons in
classically weak and quantizing magnetic fields were also
identified \cite{polyanskaya}. Thus, the experiments carried out
in 1980s showed that TQC provides a sufficiently complete
description of the low-temperature galvanomagnetic and kinetic
effects in weakly disordered two-dimensional systems. Hence, the
question about the nature of the ground state of a two-dimensional
electron system in a zero magnetic field was, for almost two
decades, considered to have a unique answer: at absolute zero of
temperature one should not expect anything but the insulating
state.

\begin{figure}\vspace{-0.8in}
\scalebox{0.45}{\includegraphics[clip]{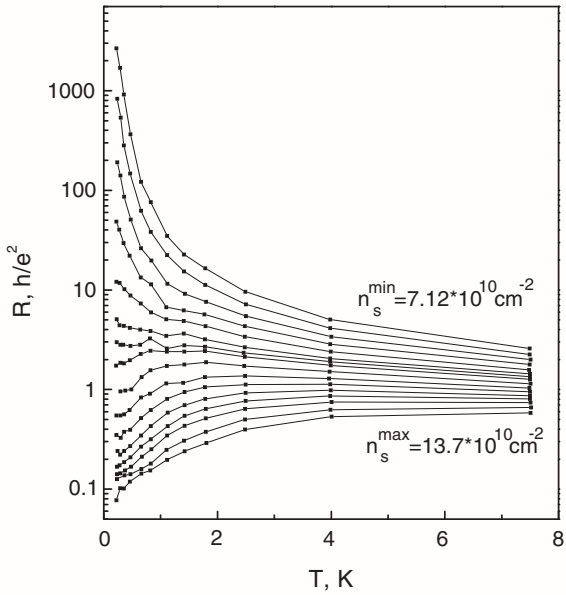}}
\vspace{-0.8in}\caption{\label{kras} Temperature dependencies of
the resistance of Si-MOSFET with a concentration of
two-dimensional electrons changing in the range
$7.12\cdot10^{10}\div13.7\cdot10^{10} $ cm$^{-2}$ in a zero
magnetic field \cite{krav1}.}\vspace{5mm}
\end{figure}

\begin{figure}\vspace{-0.8in}
\scalebox{0.45}{\includegraphics[clip]{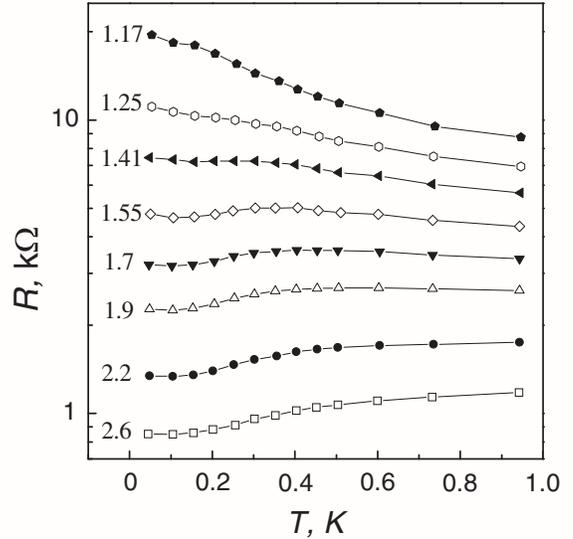}}
\vspace{-0.8in}\caption{\label{lepo} Temperature dependencies of
the resistance of $GaAs/AlGaAs$ heterostructure with a
concentration of two-dimensional holes varying in the range
$1.17\cdot10^{10}\div2.6\cdot10^{10} $ cm$^{-2}$ in a zero
magnetic field \cite{proskuryakov}.}\vspace{5mm}
\end{figure}

Against the  background of numerous experimental confirmations of
the conclusions drawn in Ref.[\cite{abrahams}] concerning the
insulating properties of the ground state of two-dimensional
systems, studies of the conductance of Si-MOSFET with a high î
(3$\cdot 10^4 $cm$^2$/(V$\cdot$s))  electron mobility
\cite{krav,krav1} had a revolutionary  character.In Refs.
[\cite{krav,krav1}], the temperature dependence of the resistance
of Si-MOSFETs with two-dimensional electron gas  in the range of
sufficiently low electron concentrations $n_s\leq 10^{11}$
cm$^{-2}$ was measured. Structures having electron concentrations
$\leq 10^{11} $cm$^{-2}$ demonstrated a usual, for localized
states, negative derivative  $dR/dT<0$ of the resistance with
respect to the temperature. However, at a certain critical
concentration  $n_c\approx 10^{11}$cm$^{-2}$, the resistance of
MOSFETs took  an approximately temperature-independent value
$R\sim 2h/e^2$.  An even more unexpected fact was a sharp decrease
of the resistance with decreasing temperature in the structures
with electron concentration  $n_s>n_c$ ,which was observed down to
the lowest experimentally reachable temperatures of ñ 200 mK in
the absence of any signs of electron localization
(Fig.~\ref{kras}). The concentration $n_c$ corresponding to a
change in the sign  of the derivative $dR/dT$, varied widely from
sample to sample, depending on the disorder in the electron system
under investigation. A similar change in the sign of the
derivative  $dR/dT$, corresponding to the critical carrier
concentration $n_c(p_c)$,  was  subsequently found in
$AlGaAs/GaAs$ heterostructures with two-dimensional electron
\cite{hanein,ribeiro} and hole
\cite{hanein1,simmons,hanein2,hamilton,yoon,simmons1,proskuryakov,hamilton1}
gas, quantum wells  $AlAs$ with two-dimensional electron gas
\cite{papadakis}, as well as quantum $SiGe$  wells with electron
\cite{lam} and hole \cite{coleridgei,leturcq} gas. However, the
temperature dependence of the resistance of these low-dimensional
systems in the temperature range $T<1 K$ turned out to be much
less than that in silicon MOS structures (Fig.~\ref{lepo}).

In the vicinity of critical carrier concentration, the resistance
of MOSFETs showed the scaling with respect to temperature:

$R(T,n_s)=f_1(|n_s-n_c|/T^{1/z\nu})$

and electric field strength

$R(E,n_s)=f_2(|n_s-n_c|/E^{1/(z+1)\nu})$

with the exponents $z=0.8\pm0.1$, and  $\nu=1.5\pm0.1$
\cite{krav1,krav2}. The product of critical indices in Si-MOSFETs
was equal to $z\nu=1.4\div1.7$ \cite{popovic}.  In experiments
studying the low-temperature transport in Si-MOSFET with varying
peak mobility, a dependence of the  $z\nu$ on the momentum
relaxation time of electrons and critical concentration  $n_c$ has
been observed \cite{pudal}. In $AlGaAs/GaAs$  heterostructures
with two-dimensional electron gas, the analysis of scaling in
temperature and electric field allowed the determination of the
critical indices $z=1.4\pm1.0$ and $\nu=1.9\pm0.9$ \cite{ribeiro}.
In $AlGaAs/GaAs$ with p-type conductance, the product of critical
indices equals to $z\nu=7.0\pm1.5$ and  $z\nu=3.8\pm0.4$ for
systems with a concentration of two-dimensional holes $p>p_c$ and
$p<p_c$, correspondingly \cite{simmons}.In $SiGe$ quantum wells
involving  two-dimensional electron gas  with a concentration
$n_s<n_c$ the product  of critical indices  is equal
$z\nu=1.6\pm0.2$ \cite{lam}. For two-dimensional hole gas in
$SiGe$ quantum wells, the values  of the product of critical
indices obtained by  different authors are equal to
$z\nu=1.6\div2$ \cite{coleridgei} and $z\nu=2.24\pm0.20$
\cite{leturcq}.

Could one consider  a transition between the regimes with
$dR/dT<0$ and $dR/dT>0$, observed at finite temperatures, as  a
manifestation of the quantum phase transition of a two-dimensional
system from the metallic to the insulating state in  a zero
magnetic field? This question should, first of all, be solved with
respect to Si-(100) MOSFET. There are a number of reasons for such
a conclusion. A change in the sign of the low-temperature
derivative $dR/dT$ at some critical concentration in Si-MOSFET
offers an experimental fact that was reliably established  by
independent groups  of researchers
\cite{krav,krav1,krav2,popovic,pudal,krav3,shakrav}.

In other systems, for example, in $AlGaAs/GaAs$ and $SiGe$ with
two-dimensional electrons or holes, the temperature dependence of
conductance on 'metallic' side of phase transition is much more
weak as compared to the Si-MOSFET (see Figs.~\ref{kras}, and
~\ref{lepo}). Further cooling of these systems results in, at
first, saturation  of some (previously 'metallic') temperature
dependencies of resistance, and then their transformation to the
regime  with $dR/dT<0$. Such an effect was observed  in
$AlGaAs/GaAs$ heterostructures with two-dimensional electron
\cite{ribeiro} and hole \cite{proskuryakov} gas, as well as for
two-dimensional holes in $SiGe$ \cite{senz} and for
two-dimensional electron gas of Si-MOSFET for vicinal orientations
of the interface between $Si$ and $SiO_2$ \cite{safonov}.

\begin{figure}\vspace{-0.8in}
\scalebox{0.45}{\includegraphics[clip]{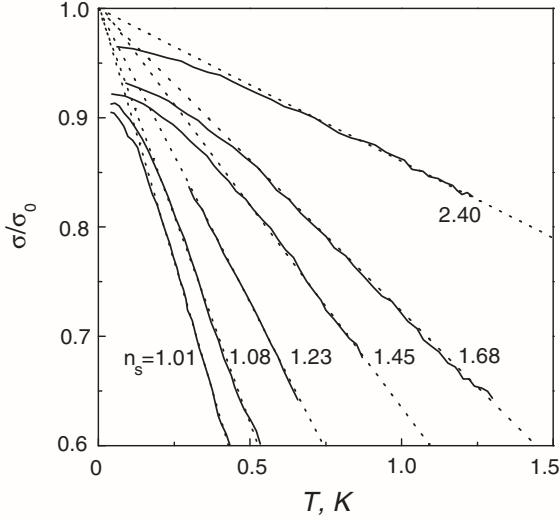}}
\vspace{-0.8in}\caption{\label{lepota} Temperature dependencies of
the normalized conductance of Si-MOSFET with the concentration of
two-dimensional electrons varying in the range
$1.01\cdot10^{11}\div2.4\cdot10^{11}$ cm$^{-2}$ in a zero magnetic
field \cite{shashkinmet}. The dashed lines show a linear
extrapolation of the temperature dependence of conductance to the
limit of absolute zero of temperature.}\vspace{5mm}
\end{figure}

\begin{figure}\vspace{-0.8in}
\scalebox{0.45}{\includegraphics[clip]{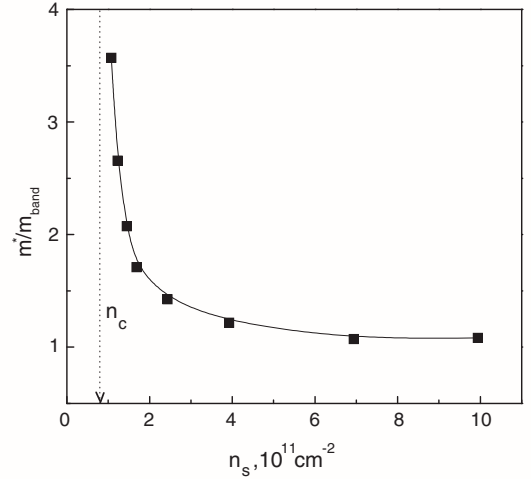}}
\vspace{-0.8in}\caption{\label{shm} Normalized effective mass as a
function of electron density in Si-MOSFET with peak mobility
$3\cdot10^4$cm$^2$/(V$\cdot$ s) \cite{shashkinmet};
$m_{band}=0.19m_e$, where $m_e$ is the mass of a free
electron.}\vspace{5mm}
\end{figure}

In numerous experiments on Si-MOSFET
\cite{krav,krav1,krav2,popovic,pudal,krav3,shakrav} it was
established that the critical carrier concentration corresponding
to the change in the sign of the derivative $dR/dT$ is determined
by the quality of the sample. Therefore, the scaling of the
dependencies $R(T)$ in the main part of investigated samples
should be considered as  being occasional \cite{maslov}. However,
an anomalously sharp growth of the conductance with decreasing
temperature cannot be  explained within classical Drude theory. A
giant change in the conductance can be caused by the change of the
screening properties of a two-dimensional system \cite{goldmet} at
a sharp reduction of the Fermi energy.

In the new experimental works  \cite{krav3,shashkinmet}, the
temperature dependence of the conductance of Si-MOSFET with the
mobility of two dimensional electrons of the order of $\sim 3*10^4
$cm$^2$/(B$\cdot$c), in which a change in the sign of the
derivative $dR/dT$ corresponds to minimal critical electron
density, was investigated in the range of ultra-low temperatures
down to  $\approx 30$ mK.  Structures  with a concentration  of
two-dimensional electrons above  the critical one  showed a rapid
linear growth of the normalized  conductance $\sigma(T)/\sigma_0$
with decreasing temperature in a sufficiently wide temperature
range (Fig.~\ref{lepota}). Interpretation of the experimentally
found linear temperature dependence  in terms of Ref.[\cite{zala}]
revealed a strong increase of the effective mass in Si-MOSFET,
when the electron density approaches the value of
$0.8\cdot10^{11}$cm$^{-2}$, which practically coincided with $n_c$
in the best studied samples \cite{shashkinmet}. Such a behavior of
the cyclotron mass was confirmed in independent experiment
\cite{pudalovmet} on the measurement of the temperature dependence
of Shubnikov -de Haas oscillations. An analysis of the
experimental data, analogous to that in Ref.[\cite{shashkinmet}],
but performed in the opposite limit with respect to the ratio of
the valley splitting energy to temperature and using evidence of
other experimental groups and samples from other sources
\cite{vit,pud},  confirmed the universality of the  $m^*(n_s)$
curve (Fig.~\ref{shm}).

As has already been mentioned, a conclusion to be made from recent
experimental findings should be that a change in the sign of the
derivative  $dR/dT$ at some concentration cannot, analogously to
the conductance scaling, be considered as convincing evidence of a
disorder driven quantum phase transition. The critical
concentration $n_c$, corresponding to a change in a sign of the
derivative $dR/dT$ varies from one sample to another. However, the
concentration  $n_c^*$ extracted from the divergence of the
effective mass most probably  takes a universal value or changes
weakly from sample to sample. For the best samples, a negative
magnetoresistance effect \cite{krav4} disappears in the vicinity
of this concentration. If a quantum phase transition in Si-MOSFET
exists, one should think that $n_c^*$ is precisely the quantum
phase transition point. Such a phase transition is a property of
pure, rather than disordered, two-dimensional systems with strong
interparticle interactions, in particular, of the most perfect
MOSFETs with a low electron concentration \cite{abr}.

\section{Conclusions}

In many experimental studies of the quantum phase transitions in
two-dimensional systems, the emphasis is, in our opinion, put on
the facts that confirm the theory. The facts that are difficult to
interpret within theoretical schemes are silently ignored. The
full phase diagrams, similar to that shown in Fig.~\ref{QPT}, have
been, to the best of our knowledge, for the first time constructed
from experimental data only in the present review (see
Figs.~\ref{super2},~\ref{super4},~\ref{phase}). A law describing
the boundary of the quantum critical region can be determined from
experiment and exploited for an independent control of the scaling
relations. This possibility of independent control was, however,
never used in the analysis of experimental data.

Therefore, existing theoretical ideas that successfully predicted
and explained a number of experimental observations can hardly be
currently considered as having been confirmed experimentally.

The authors express their sincere gratitude to V.F. Gantmakher,
S.N. Molotkov, and V.V. Ryazanov  for many valuable comments. The
work  was supported  by RFBR, RF Ministry of Industry, Science and
Technology and State Program of Support of Scientific Schools.

\end{document}